\documentclass[preprint,12pt]{elsarticle}

\usepackage{amssymb}
\usepackage{amsmath}
\usepackage{url,moreverb,xspace}
\usepackage{booktabs} 
\usepackage[flushleft]{threeparttable}
\usepackage{xcolor}
\usepackage[utf8]{inputenc}
\usepackage{float}
\usepackage{amssymb}
\usepackage{amsmath}
\usepackage{ifthen}
\usepackage{caption}
\usepackage[bookmarks=false]{hyperref}
\usepackage{url,moreverb,xspace}
\usepackage{tcolorbox}
\usepackage{enumitem}
\usepackage{array,graphicx}
\usepackage{soul}
\usepackage{balance}
\usepackage{pifont}
\usepackage{nicefrac}
\usepackage{mathtools}
\usepackage{tabularx}
\usepackage{xcolor,pifont}
\usepackage{tikz}
\usepackage{svg}
\usepackage{pdfpages}
\usepackage{array}
\usepackage{subcaption}
\usepackage{microtype}

\newcommand*\colourcheck[1]{%
  \expandafter\newcommand\csname #1check\endcsname{\textcolor{#1}{\ding{52}}}%
}
\newcommand*\colourcross[1]{%
  \expandafter\newcommand\csname #1cross\endcsname{\textcolor{#1}{\ding{56}}}%
}
\usepackage[vlined, boxruled, linesnumbered] {algorithm2e}
\SetKw{KwBy}{by}
\SetKw{KwBreak}{break}
\SetKw{KwReturn}{return}

\newcommand{\tool}{\mbox{PerturbationDrive}\xspace}

\newboolean{showcomments}
\setboolean{showcomments}{true}
\ifthenelse{\boolean{showcomments}}
{\newcommand{\nb}[2] {
  \fcolorbox{black}{gray!20}{\bfseries\sffamily\scriptsize#1:}
  {\sf\small$\blacktriangleright$\textit{#2}$\blacktriangleleft$}
}
}
{\newcommand{\nb}[2]{}
}

\journal{Science of Computer Programming}

\begin{document}

\begin{frontmatter}



\title{PerturbationDrive: A Framework for Perturbation-Based Testing of ADAS}




\author{Hannes Leonhard}
\address{Technical University of Munich, Germany}

\author{Stefano Carlo Lambertenghi}
\address{Technical University of Munich \& fortiss, Germany}

\author{Andrea Stocco}
\address{Technical University of Munich \& fortiss, Germany}

\begin{abstract}
Advanced driver assistance systems (ADAS) often rely on deep neural networks to interpret driving images and support vehicle control. Although reliable under nominal conditions, these systems remain vulnerable to input variations and out-of-distribution data, which can lead to unsafe behavior. To this aim, this tool paper presents the architecture and functioning of \tool, a testing framework to perform robustness and generalization testing of ADAS. The framework features more than 30 image perturbations from the literature that mimic changes in weather, lighting, or sensor quality and extends them with dynamic and attention-based variants. \tool supports both offline evaluation on static datasets and online closed-loop testing in different simulators. Additionally, the framework integrates with procedural road generation and search-based testing, enabling systematic exploration of diverse road topologies combined with image perturbations.  
Together, these features allow \tool to evaluate robustness and generalization capabilities of ADAS across varying scenarios, making it a reproducible and extensible framework for systematic system-level testing.
\end{abstract}




\end{frontmatter}


\section*{Metadata} \label{sec:Meta}

\begin{table}[H]
\caption{Code metadata}
\label{tab:code_metadata}
\begin{tabular}{|p{6.5cm}|p{6.5cm}|}
\hline
\textbf{Code metadata description} & \textbf{Please fill in this column} \\
\hline
Current code version & v1.0.0 \\
\hline
Permanent link to code/repository used for this code version & \url{https://github.com/ast-fortiss-tum/perturbation-drive.git} \\
\hline
Legal Code License & MIT \\
\hline
Code versioning system used & git \\
\hline
Software code languages, tools, and services used & Python 3.9, Unity (C\#), pygame \\
\hline
Compilation requirements, operating environments and dependencies &  Unix (x86/arm)\\
\hline
Link to developer documentation/manual & \url{https://github.com/ast-fortiss-tum/perturbation-drive/blob/Replication/README.md} \\
\hline
Support email for questions & \href{lambertenghi@fortiss.org}{lambertenghi@fortiss.org} \\
\hline
\end{tabular}
\end{table}
\section{Introduction}\label{sec:intro}

Advanced driver assistance systems (ADAS) use perception modules to interpret driving environments in real time for tasks such as object detection, segmentation, and control regression~\cite{survey-lei-ma,yurtsever2020survey,li2024panopticperceptionautonomousdriving,grigorescu2020survey,2026-Guo-OJ-ITS,RiccioEMSE20}. Deep neural networks (DNNs) represent the standard methodology for ADAS perception and currently deliver the best reported performance. 
Although accurate under nominal conditions, DNNs often fail to generalize to unseen domains. Exhaustive data collection is infeasible, and small shifts in lighting, weather, or viewpoint can cause perception errors~\cite{li2024panopticperceptionautonomousdriving,dodge2016understanding,geirhos2020generalisation,2020-Humbatova-ICSE,2026-Naziri-ICSE,2020-Stocco-ICSE} that can produce unsafe driving behaviors.

Model-level testing provides insights but ignores the closed-loop nature of driving, where perception continuously affects control~\cite{2023-Stocco-EMSE}. Large-scale validation requires simulators, since real-world testing is unsafe and would demand millions of miles~\cite{10-million-miles,2025-Lambertenghi-ASE}. Platforms such as CARLA~\cite{carla}, Udacity~\cite{udacity-sim}, DonkeyCar~\cite{donkeycar}, and NVIDIA DriveSim~\cite{nvidia2023drivesim} support reproducible evaluation, but realistic adverse conditions often require custom assets. Perturbation-based methods address this gap by directly manipulating images, ensuring portability across datasets and simulators.

In our previous work, we presented \tool~\cite{2025-Lambertenghi-ICST}, a library that integrates several perturbations from the literature for vision-based ADAS testing~\cite{hendrycks2019benchmarking,hendrycks2020augmix,rusak2020simple,Laermann-2019,8388338,2023-Stocco-TSE,ayerdi2023metamorphic}. In this work, we extended \tool in several directions. First, we implemented dynamic~\cite{10208325} and attention-guided variants~\cite{10043233,Kitada2020AttentionMP} of the original perturbations. Additionally, we integrate perturbations into search-based testing for combined scenario–perturbation exploration~\cite{9476240,HUMENIUK2023102990,sorokin2025simulatorensemblestrustworthyautonomous,sorokin2023opensbt} and we added support to the CARLA simulator. \tool supports both offline (component-level)~\cite{deepxplore,deeptest,deepbillboard} and online (system-level) evaluation~\cite{2023-Stocco-EMSE,dataAugment2020Liu,yoon2023learning,2024-Grewal-ICST} to enable systematic and reproducible evaluation of ADAS, covering both robustness and generalization across diverse driving scenarios. 

\section{The \tool Framework}
\label{sec:framework}

\subsection{Objectives}

The goal of \tool is to provide a systematic framework for evaluating ADAS under controlled image perturbations and procedurally generated road scenarios. It consolidates perturbation techniques into a library with standardized configurations for type, intensity, and random seed, ensuring reproducibility and comparability across ADAS models and simulation environments. Beyond offline testing, \tool supports closed-loop evaluation in simulators and integrates perturbations with procedural road generation, enabling both robustness benchmarking and generalization analysis in unseen conditions.  

\subsection{System Architecture}

\begin{figure*}[t]
  \centering
    \includegraphics[width=\linewidth]{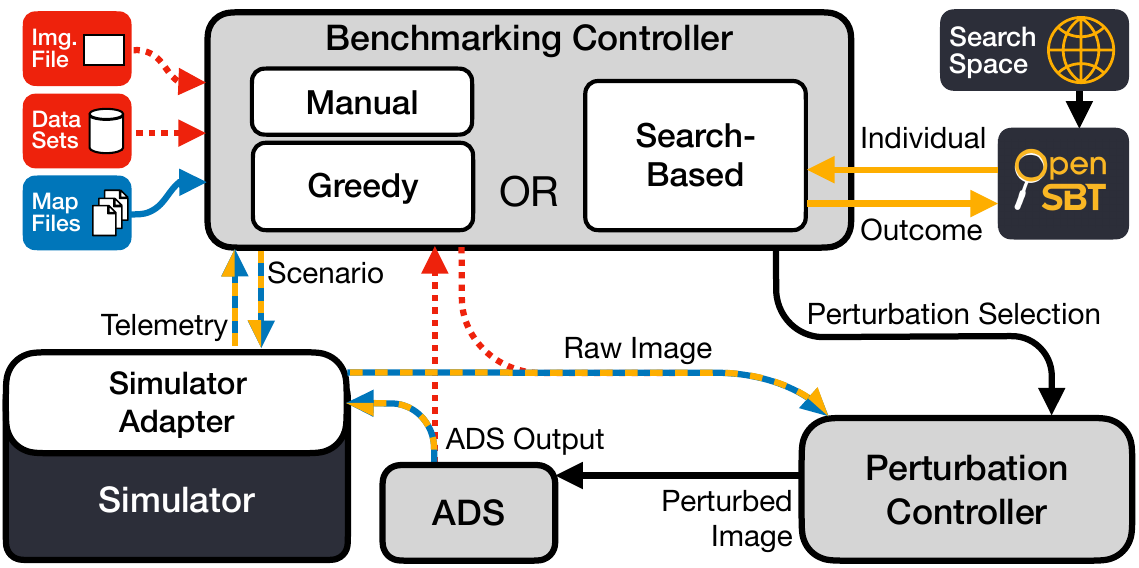}
    \caption{Overview of the \tool framework.\\\textbf{Red:} Offline execution, \textbf{Blue:} Online Greedy execution,  \textbf{Yellow:} Online SBT execution.}
    \label{fig:framework}
\end{figure*}

The framework comprises three components (\autoref{fig:framework}): the \textit{Perturbation Controller}, the \textit{Simulator Adapter}, and the \textit{Benchmarking Controller}.

\subsubsection{Perturbation Controller}
The Perturbation Controller implements a library of image perturbations available from the literature~\cite{hendrycks2019benchmarking,hendrycks2020augmix,rusak2020simple,Laermann-2019,8388338,2023-Stocco-TSE,ayerdi2023metamorphic}. Perturbations are grouped into three categories:\\
\textbf{i)~static perturbations} include frame-level modifications such as noise, blur, defocus, weather overlays, geometric distortions, affine transformations, graphic patterns, and color or tone adjustments.\\  
\textbf{ii)~dynamic perturbations} propose temporal overlays (e.g., rain, snow, smoke, glare) that preserve consistency across frames.\\  
\textbf{iii)~attention-guided perturbations}: perform targeted distortions applied to salient regions identified by GradCAM or similar methods~\cite{2025-Chen-EMSE,2022-Stocco-ASE}. All perturbations inherit from a common base class that defines the transformation interface for consistency and extensibility.  

\subsubsection{Simulator Adapter}

The Simulator Adapter bridges the perturbation library and simulators. It intercepts raw camera frames, applies perturbations, and forwards modified frames to the ADAS under test. Current implementations support Udacity~\cite{udacity-simulator}, DonkeyCar~\cite{donkey}, and CARLA~\cite{carla}; additional platforms (e.g., BeamNG~\cite{gambi-beamng}, NVIDIA DriveSim~\cite{nvidia2023drivesim}) can be added by implementing the adapter interface. The adapter also integrates with procedural road generation to produce diverse topologies and enforces a per-frame processing budget to maintain real-time execution at 30 FPS.  

\subsubsection{Benchmarking Controller}

The Benchmarking Controller manages experiments in two main modes: \textit{offline} and \textit{online}~\cite{2023-Stocco-EMSE}. In offline mode, datasets are perturbed, and the resulting model outputs are compared against ground truth or reference predictions. In online mode, the controller supervises closed-loop simulations, injecting perturbations in real time while logging frames, control actions, and vehicle states, and detecting failures such as collisions or lane departures. During online execution, all parameters and execution traces are recorded to ensure reproducibility and enable replay of failure-inducing cases.\\
For both offline and online experiments, perturbations can be applied either manually or through a greedy strategy. Additionally, for online experiments, \tool supports a search-based testing (SBT) approach to guide the exploration of perturbations and scenarios.\\
\textbf{Greedy Testing.} Each perturbation is applied to the provided dataset or road scenarios, starting from a low intensity. In offline experiments, all perturbations are evaluated across all intensity levels. In online experiments, if a system failure occurs, the controller proceeds to the next perturbation; otherwise, the perturbation intensity is progressively increased until a failure occurs or all intensity levels have been tested.\\
\textbf{Search-Based Testing.} The combined space of perturbation types, intensities, and road scenarios is too large for exhaustive evaluation. To address this, \tool incorporates SBT. Perturbations and scenarios are treated as search parameters, and candidate tests are generated using fitness functions computed from simulation outcomes to prioritize combinations most likely to expose system failures. In our implementation, we leverage OpenSBT~\cite{sorokin2023opensbt} to guide this search process. Treating perturbations as first-class search variables enables unified exploration of environmental diversity and perceptual distortions, moving robustness evaluation beyond ad hoc perturbation studies toward the systematic discovery of safety-critical failures in ADAS systems.

\subsection{APIs and Modularity}

The user-facing API supports three modes:\\  
\textbf{i)~image-level} perturbs a single image for visualization or debugging.\\ 
\textbf{ii)~dataset-level} perturbs entire datasets to benchmark classifiers, detectors, or segmentation models.\\  
\textbf{iii)~online} perturbs live simulator streams for end-to-end system evaluation.  

Users specify perturbation type and intensity in all modes. The framework records parameters and random seeds to ensure consistent application. Modularity is achieved by separating concerns: the Perturbation Controller defines transformations, the Simulator Adapter ensures simulator-agnostic integration, and the Benchmarking Controller handles logging and execution. This layered design facilitates extension with new perturbations or simulators without altering existing code. 

\section{Implementation}\label{sec:implementation}

\subsection{Static Perturbations}
\label{sec:list}

\tool provides 38 static perturbations, grouped into eight categories:\\
A)~Noise perturbations, which mimic sensor or compression artifacts, including Gaussian, Poisson, impulse (salt-and-pepper), JPEG, and speckle noise.\\
B)~Blur and focus perturbations, which reduce image sharpness through defocus, motion, zoom, Gaussian, or low-pass blur.\\
C)~Weather perturbations, which simulate adverse conditions such as frosted glass, snow, fog, brightness shifts, and contrast changes.\\
D)~Distortion perturbations, which deform spatial structure via elastic deformation, pixelation, region blending, or sharpening.\\
E)~Affine transformations, which alter global geometry through shear, scaling, translation, rotation, or reflection.\\
F)~Graphic pattern perturbations, which overlay artificial structures such as splatter, dotted lines, zig-zags, edge maps, or cutout masks.\\
G)~Color and tone adjustments, which change appearance by applying false colors, scrambling, histogram equalization, white balance, greyscale, saturation, or posterization.\\
H)~Generative perturbations, which use deep models such as CycleGAN for domain remapping or style transfer for artistic overlays.

\begin{figure*}[t]
  \centering
  \begin{subfigure}{0.98\textwidth}
    \includegraphics[width=\linewidth]{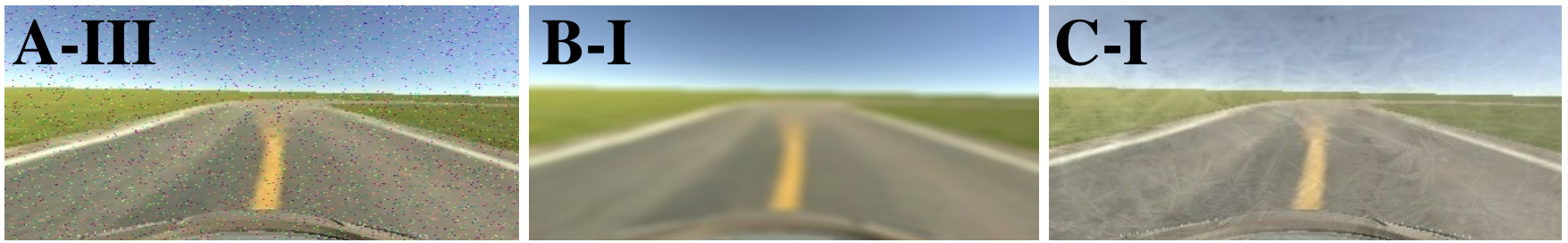}
    \caption{Valid.}
  \end{subfigure}
  \hspace{0.01\textwidth}
  \begin{subfigure}{0.98\textwidth}
    \includegraphics[width=\linewidth]{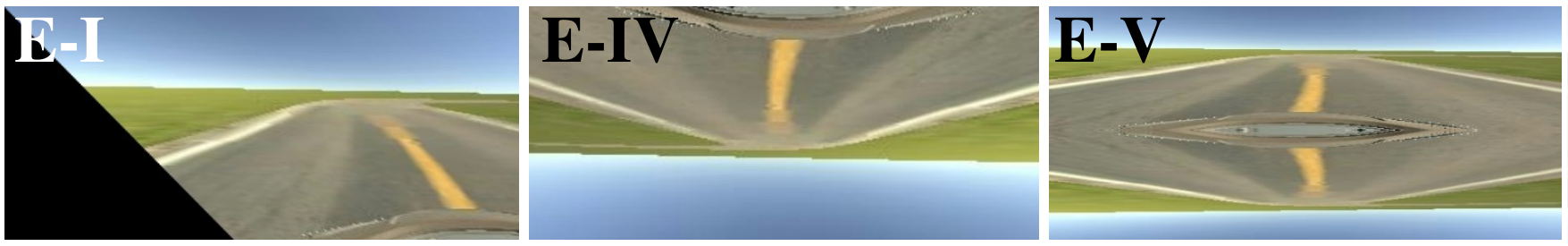}
    \caption{Invalid.}
  \end{subfigure}
  \caption{Examples of valid and invalid static perturbation types.}
  \label{fig:invalid-perturbations}
\end{figure*}

To ensure meaningful robustness evaluation, we manually inspected each perturbation and selected a standard set that preserves scene semantics and produces valid driving images (\autoref{fig:invalid-perturbations}a). Transformations that distort the scene unrealistically (e.g., removing the road, vertical flips)~were excluded from the default configuration (\autoref{fig:invalid-perturbations}b).  

\subsubsection{Intensity Levels}

Perturbation intensity was calibrated by visual inspection. For static perturbations, the strength of the effect was increased gradually until the scene could no longer be reliably interpreted by a human observer (i.e., a human could no longer confidently infer an appropriate driving action from the image). Three assessors performed this calibration independently, and the final threshold was determined by their agreement. The maximum valid intensity was set just below this threshold, and the resulting range was divided into five uniform intensity levels.

For dynamic perturbations, intensity is controlled through the transparency parameter $\alpha$. The maximum intensity corresponds to full visibility of the perturbation ($\alpha=1.0$), while the minimum intensity corresponds to 80\% transparency ($\alpha=0.20$). This interval was then uniformly divided into five intensity levels.

\subsection{Dynamic Perturbations}
Dynamic perturbations in \tool are implemented in two forms: video overlays and particle-based effects. These approaches maintain temporal consistency and remain simulator-agnostic since overlays are applied at the frame level rather than through engine-specific weather models.  

\subsubsection{Overlay-based effects} 

\begin{figure*}[t]
  \centering
  \begin{subfigure}{0.46\textwidth}
    \includegraphics[width=\linewidth]{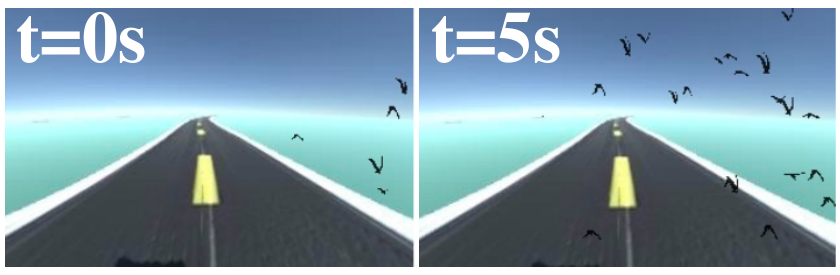}
    \caption{Overlay.}
  \end{subfigure}
  \hspace{0.01\textwidth}
  \begin{subfigure}{0.50\textwidth}
    \includegraphics[width=\linewidth]{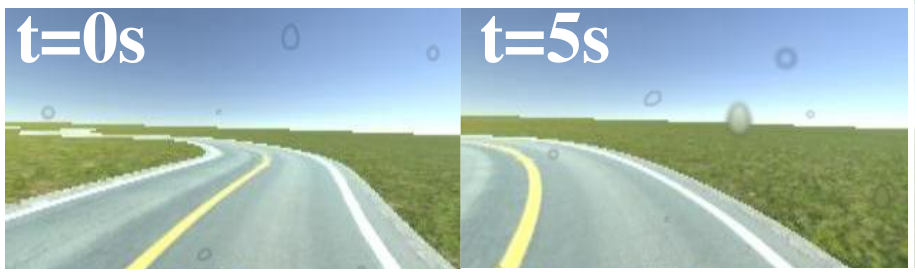}
    \caption{Particle.}
  \end{subfigure}
  \caption{Examples of overlay-based and Particle-based dynamic perturbation types.}
  \label{fig:dynamic}
\end{figure*}

Each effect (e.g., rain streaks, snow, smoke, birds, glare)~is represented by a pre-recorded green-screen video clip. During evaluation, chroma-key removal sets green pixels to transparent, blending only the perturbation elements into the scene. Temporal consistency is ensured using a \texttt{CircularBuffer}, which aligns simulator frames with the correct overlay frame, preserving natural motion such as continuous snowfall or rain streaks. Users may also supply custom overlays without coding: any green-screen video can be automatically processed and injected into the simulation.\\
An example spanning five seconds of simulation is shown in \autoref{fig:dynamic}a.

\subsubsection{Particle-based effects}

In addition to overlays, \tool implements perturbations inspired by how raindrops or snowflakes interact with a physical camera lens. When precipitation strikes the lens, droplets or flakes attach to the glass, slowly accumulate, and then drift across the field of view under gravity and airflow. This creates localized occlusions that move over time, degrading visibility in a way that global weather overlays cannot reproduce.  
In \tool, these effects are simulated by representing each droplet or flake as a particle with position, size, and velocity updated stochastically at every frame. Initial positions are sampled randomly (or from salient regions in the attention-guided variant), and their trajectories evolve according to random lateral drift, downward motion, and size adjustments. A \texttt{CircularBuffer} is used to maintain temporal consistency, ensuring that droplets and flakes persist and move smoothly across frames instead of flickering. For rain, particles appear as semi-transparent streaks that may merge or slide, mimicking water on glass. For snow, flakes fall more slowly with wider drift, producing accumulation-like patterns.  
This particle-based design enables realistic simulation of precipitation on camera lenses, complementing static and overlay-based perturbations. 
An example spanning five seconds of simulation is shown in \autoref{fig:dynamic}b.


\begin{figure*}[t]
  \centering
  \begin{subfigure}{0.98\textwidth}
    \includegraphics[width=\linewidth]{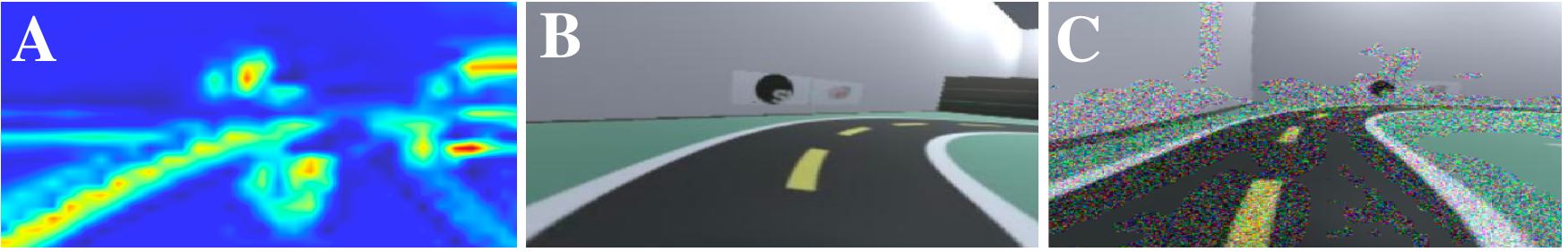}
  \end{subfigure}
  \caption{Example of attention-based perturbation.}
  \label{fig:attention}
\end{figure*}

\subsection{Attention-Based Perturbations}
Attention-based perturbations in \tool apply changes only to the parts of the image that the system under test considers important~\cite{2025-Chen-EMSE,2022-Stocco-ASE}.

The process begins with saliency extraction. By default, \tool uses GradCAM to create a saliency map, which highlights the pixels the DNN relies on most. The map is resized to the input resolution and normalized to values between $0$ and $1$. Other attribution methods can be substituted as long as they return a pixel-level saliency map.  

From this saliency map, a mask is generated. Pixels can be selected in several ways: keeping all values above a fixed threshold $\epsilon$, selecting the top-$n$\% of values (high-saliency regions such as lane markings or vehicles), or selecting the bottom-$n$\% (low-saliency regions such as sky or trees). A random baseline can also be used, where the same number of pixels is chosen at random.  

To ensure valid masks, small scattered areas are removed, and optional morphological closing is applied to form continuous regions. Soft masks with alpha blending can also be used, allowing a gradual rather than sharp boundary. In practice, for ADAS, background regions and ground outside the drivable surface are usually not relevant, so masks are focused on road areas and traffic participants.  

Any static perturbation from the library (e.g., Gaussian noise, blur, occlusion, brightness change)~can then be applied to the pixels within the mask. The final perturbed image is obtained by combining the perturbed pixels with the original image using the mask: pixels inside the mask are replaced by their perturbed versions, while pixels outside the mask remain unchanged. This ensures that only the masked areas are modified and the rest of the image stays intact.
Saliency maps can be cached for offline evaluation or recomputed periodically during online testing. Masks may also be reused across short horizons to meet real-time constraints.
An example of an attention-based perturbation is shown in \autoref{fig:attention} (A: saliency map, B: original image, C: Perturbed image).

In addition to static masking, \tool extends attention-based perturbations with dynamic precipitation effects guided by saliency. Prior work, such as AdvRain~\cite{info14120634}, has shown that placing synthetic raindrops at salient regions can create adversarial perturbations that mislead vision systems. Our approach differs in that the precipitation is not static: droplets are generated at salient regions on the lens and then drift until leaving the field of view, while snowflakes are emitted from salient regions and repeatedly cover important areas.

\subsection{Perturbations in Search-based Testing}

Each test case is represented as a tuple (\text{road scenario}, \text{perturbation type}, \text{intensity}). Road scenarios are generated procedurally, while perturbations are applied to the camera stream.  

A key requirement is an ordering of perturbations by effect strength; otherwise, type would be categorical, complicating evolutionary search. To address this, we established a ranking based on an empirical study across 2,450 scenarios (49 perturbations × 5 intensity levels × 10 roads)~using the Udacity and DonkeyCar simulators, measuring degradation of an end-to-end lane-keeping and adaptive cruise control model, DAVE-2~\cite{nvidia-dave2}. 

The ten roads were procedurally generated using the road generation functionality provided by the simulator adapter by specifying sequences of eight waypoint angles and fixed segment lengths. Each road consists of eight consecutive segments of 25\,m, where the curvature is controlled by the relative angle between adjacent waypoints. The angles range from $0^\circ$ for straight segments to $\pm 35^\circ$ for sharp turns, producing roads with varying combinations of straight sections, moderate curves, and high-curvature bends. To ensure valid benchmarking conditions, each generated road was verified to be fully drivable by the ADS without perturbations, confirming that the system could successfully complete all ten roads under nominal conditions.
This ranking allows perturbation type to be treated as an ordinal parameter in the SBST search space. The Benchmarking Controller integrates this ordering with the search-based testing framework OpenSBT~\cite{sorokin2023opensbt}, using the integer variant of the NSGAII-DT~\cite{ALGO} evolutionary algorithm. Candidate scenarios are represented by four parameters: the number of road turns, the average road curvature, the perturbation intensity level, and the perturbation type, mapped to an ordinal variable based on the empirical ranking described above.

During each iteration, OpenSBT generates candidate test cases by sampling this search space and executing them in simulation. The fitness function is defined as a four dimensional objective vector consisting of: (i)~the average absolute cross track error (XTE), which measures the lateral distance of the vehicle from the center of the driving lane, (ii)~the time to failure, (iii)~a criticality indicator derived from simulation outcomes such as timeouts or unsuccessful scenario completion, and (iv)~the maximum observed XTE. The search aims to maximize the average XTE, time to failure, and maximum XTE, while minimizing the criticality indicator. All metrics are computed by the Benchmarking Controller using telemetry data provided by the simulator adapter.

The evolutionary search is configured by default with a population size of 75 and runs for 50 generations, resulting in up to 7,500 fitness evaluations. Mutation uses a distribution index of $\eta = 20$. To ensure reproducibility, the random seed is fixed, which controls both the initialization of the population and the stochastic components of scenario generation. Each candidate scenario is simulated for 30 seconds with a sampling interval of 0.25 seconds. All of these parameters can be changed by the user to better fit experimental needs.

\section{Usage}
\label{sec:usage}

\subsection{Installation}
\tool is available as a Python package, which supports Python $>$= 3.9 (although Python 3.9 is the preferred version).
For up-to-date instructions, please refer to the tool's repository on GitHub.

We recommend using Micromamba~\cite{Micromamba_2026} to create a virtual environment to ensure the correct Python version and  tkinter~\cite{Tkinter_2026} support, which is a required component.
To install Micromamba:
\begin{verbatim}
curl -Ls https://micro.mamba.pm/install.sh | bash
source ~/.bashrc
\end{verbatim}
Then create a virtual environment:
\begin{verbatim}
micromamba create -n myenv python=3.9
\end{verbatim}
If necessary, install the appropriate Python version:
\begin{verbatim}
sudo apt update
sudo apt install software-properties-common
sudo add-apt-repository ppa:deadsnakes/ppa
sudo apt update
sudo apt install python3.9 python3.9-venv python3.9-distutils
\end{verbatim}
Then activate the environment:
\begin{verbatim}
micromamba activate myenv
\end{verbatim}
Clone the repository and install dependencies. Please replace \textbf{[operating\_system]} with your Operating System; currently we provide requirements for linux with X86 architectures \textbf{[linux]} and MacOS with Apple Silicon \textbf{[macos]}:
\begin{verbatim}
git clone https://github.com/ast-fortiss-tum/perturbation-drive.git
cd perturbation-drive
pip install -r requirements_[operating_system].txt
\end{verbatim}
In case the requirements installation fails, we provide an alternative file:
\begin{verbatim}
pip install -r requirements_other.txt
\end{verbatim}
Install the library locally:
\begin{verbatim}
pip install .
\end{verbatim}
Verify the installation with the minimal example:
\begin{verbatim}
python test_standalone_perturbations.py
\end{verbatim}

The library can be used by importing perturbation functions through the \texttt{ImagePerturbation} manager, or via the included simulator test scripts. 

\subsection{Offline Evaluation}

\paragraph{Direct function calls}  
All perturbations are available as Python functions that take an image and an intensity parameter. For example:
\begin{verbatim}
from perturbationdrive import gaussian_noise, fog_filter
import cv2

image = cv2.imread("image.png", cv2.IMREAD_UNCHANGED)
perturbed = gaussian_noise(3, image.copy())   # intensity = 3
cv2.imwrite("gaussian.png", perturbed)
\end{verbatim}
This approach is suitable for visualizing or debugging single perturbations.  
A minimal example script is provided: \texttt{test\_standalone\_perturbations.py}.

\paragraph{Manager-based interface}  
Multiple perturbations can also be applied using the \texttt{ImagePerturbation} class, which dispatches calls by perturbation name:
\begin{verbatim}
from perturbationdrive import ImagePerturbation
import cv2

image = cv2.imread("0001.png", cv2.IMREAD_UNCHANGED)
perturbation_names = ["gaussian_noise", "snow_filter"]

controller = ImagePerturbation(funcs=perturbation_names)

for p in perturbation_names:
    out = controller.perturbation(image.copy(), p, intensity=2)
    cv2.imwrite(f"0001_{p}.png", out)
\end{verbatim}
A minimal example script is provided: \texttt{test\_perturbation\_manager.py}.

\subsection{Dataset Evaluation}
Dataset evaluations can be performed using the minimal example script
\texttt{test\_dataset.py}. The script iterates over all images in a dataset and
applies different perturbations with varying intensities. For each unique
combination of image, perturbation type, and perturbation intensity, the model
generates driving commands for both the original and the perturbed image.
These commands are stored together with the corresponding ground truth actions
for later evaluation in a JSON file.

The dataset folder must contain two types of files:
\begin{itemize}
  \item Image frames named in the format\\
  \texttt{frameNumber\_freeString.\{jpg|jpeg|png\}}, where
  \texttt{freeString} is an arbitrary string of length $>1$.
  \item A JSON file for each frame named
  \texttt{record\_frameNumber.json}, where \texttt{frameNumber}
  corresponds to the associated image.
\end{itemize}

Each JSON file must contain the ground truth driving commands:
\begin{itemize}
  \item Steering angle stored in \texttt{user/angle}
  \item Throttle stored in \texttt{user/throttle}
\end{itemize}
\subsection{Online Evaluation}
Closed-loop perturbations in simulators are demonstrated by two minimal example scripts:  

\begin{itemize}
  \item \texttt{test\_sim\_udacity.py} for the Udacity simulator,  
  \item \texttt{test\_sim\_donkey.py} for DonkeyCar.  
\end{itemize}

Both scripts connect to the simulator, intercept camera frames, apply perturbations, and feed them to the model under test.
The perturbation type and intensity are defined inside each script and can be modified by editing the corresponding calls to \texttt{ImagePerturbation}.
A CARLA simulator adapter is provided in \texttt{examples/carla/}.

\subsection{Extending the Library}
\textbf{Adding a perturbation.}  
New perturbations can be added by defining a function in \texttt{perturbationdrive/} that follows the interface \texttt{func(intensity, image)}. They can also be registered for use with \texttt{ImagePerturbation}.  

\textbf{Adding a simulator.}  
Additional simulators can be integrated by following the structure of \texttt{test\_sim\_udacity.py} and \texttt{test\_sim\_donkey.py}, where perturbations are injected into the frame-processing loop.

\section{Expected Impact and Significance} \label{sec:imp}

PerturbationDrive provides a complete experimental environment for assessing the reliability of ADAS and the interplay between robustness and generalizability under controlled conditions.
It supports systematic comparisons of image perturbations and road topologies across diverse ADAS and simulation environments. Its automated workflow minimizes manual coding effort while ensuring reproducibility and extensibility. We hope that the tool will serve as an accelerator for researchers, students, and practitioners, enabling them to conduct reproducible robustness evaluations through intuitive interfaces that lower entry barriers and facilitate both experimentation and analysis.
\section{Conclusions and Future Work} \label{sec:con}

We presented PerturbationDrive, a framework that combines image perturbations with evolutionary algorithms to systematically assess the quality of DNN-based ADAS. By integrating existing perturbations within a reusable framework, the tool enables automated and reproducible testing across multiple simulators. 
Future work will focus on enhancing the flexibility and extensibility of the framework, enabling seamless integration of custom ADAS, datasets, and generative perturbations~\cite{2025-Baresi-ICSE,2024-Lambertenghi-ICST}, thus broadening its applicability for both research and practice. 

\section{Acknowledgments}\label{sec:acks}

This work was supported by the Bavarian Ministry of Economic Affairs, Regional Development, and Energy.




\bibliographystyle{elsarticle-num} \bibliography{paper.bib}



\end{document}